\documentclass[twocolumn,showpacs,preprintnumbers,amsmath,amssymb]{revtex4-2} 
\usepackage{graphicx}
\usepackage{dcolumn}
\usepackage{bm}
\usepackage{tabularx}
\usepackage{ulem} 
\newcolumntype{M}{>{\centering\arraybackslash}m{1.85cm}}
\usepackage[export]{adjustbox}
\usepackage{float}
\usepackage{braket}
\usepackage{lipsum}
\usepackage{amsmath}
\graphicspath{{figure/}}
\usepackage{xcolor}   

\usepackage{caption}
\usepackage{subcaption}

\usepackage{hyperref}
\hypersetup{
    colorlinks=true, 
    linktoc=all,     
    urlcolor  = cyan,
    citecolor = blue,
    linkcolor=blue,  
}

\makeatletter
\newcommand{\colorcaption}[2][]{%
  \begingroup%
  \renewcommand{\@caption@fignum@sep}{ (Color online). }%
  \caption[#1]{#2}%
  \endgroup%
}
\makeatother
\newcommand{\orcid}[1]{\href{https://orcid.org/#1}{\hskip2pt\includegraphics[width=9pt]{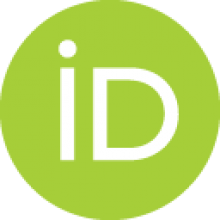}}}

\begin{document}

\title{ Evolution of shell structure at $\mathbf{N=32}$ and 34: Insights from realistic nuclear forces }

\author{Subhrajit Sahoo\orcid{0000-0001-8000-2150}}
\email{s$\_$sahoo@ph.iitr.ac.in}
\affiliation{Department of Physics, Indian Institute of Technology Roorkee, Roorkee 247667, India}

\author{Praveen C. Srivastava\orcid{0000-0001-8719-1548}}
\email{ praveen.srivastava@ph.iitr.ac.in}
\affiliation{Department of Physics, Indian Institute of Technology Roorkee, Roorkee 247667, India}

\date{\hfill \today}

\begin{abstract}
We investigated the evolution of shell structure at $N=32$ and 34 in neutron-rich nuclei beyond the stability line using realistic nuclear forces, employing the state-of-the-art valence-space in-medium similarity renormalization group method. The shell gaps are discussed from the excitation energies of the first $2^+$ states and the evolution of effective single-particle energies. We addressed different components of the nuclear interaction--central, spin-orbit, and tensor--and their roles in the development of shell gaps far from stability. The calculated results align well with the available experimental data and suggest a strengthening of the $N=34$ subshell gap and a weakening of the $N=32$ subshell gap below Ca. Additionally, the low-energy structures of the exotic $N=32$ isotones below Ca revealed that their ground states exhibit large deformation and coexist with a weakly deformed band at low excitation energy. The present work demonstrates essential components of the nuclear force in shaping magic numbers far from stability and provides deeper insights into the structure of exotic nuclei from the underlying nuclear forces.  
\end{abstract}


\maketitle
\textit{Introduction.} Nuclei, characterized by large $N/Z$ ratios, often referred to as exotic nuclei, challenge the traditional understanding of nuclear shell structure. The canonical magic numbers, such as $N=8$, 20, 28, 40, and 50, which are believed to be immutable, tend to diminish or vanish, and new magic numbers such as $N=14$, 16, 32, and 34 emerge in exotic systems \cite{SorlinReview, OtsukaReview}. In the past years, the appearance of new magic numbers, $N=32$ and 34, in the vicinity of doubly magic $^{48}$Ca has drawn significant interest from both experimental and theoretical studies \cite{OtsukaReview, Hagen2012, Hergert2014, Soma2021, Steppenbeck2013, Xu2019, Leistenschneider2021, Iimura2023, Steppenbeck2015, Rosenbusch2015, Liu2019, Cortes2020, Linh2024}. Experimental observations established $N=32$ subshell closure in neutron-rich Ti to Ca isotopes and the onset of a new shell gap at $N=34$ in Ca \cite{Steppenbeck2013, Xu2019, Leistenschneider2021, Iimura2023}. Additionally, further investigations revealed the persistence of the $N=34$ subshell gap in Ar, extending even below Ca \cite{Linh2024}. This region of the Segrè chart provides valuable inputs for understanding the shell evolution and nature of nuclear interactions away from the stability line.

The magicity at $N=32$ and $N=34$ in Ca isotopes was first predicted by Otsuka \textit{et al.}, as a consequence of strong, attractive tensor force between the $\pi f_{7/2}$ and $\nu f_{5/2}$ orbitals \cite{OtsukaPRL2001}. Subsequent studies using various effective interactions supported the emergence of a new magic number at $N=34$ in Ca \cite{ OtsukaPRL2005, Honma2002, Honma2004, Honma2005, OtsukaPRL2010, BhoyCa2020, UtsunoReview2022}, which was later confirmed by Steppenbeck \textit{et al.} through measurements of the $2^+$ excitation energy in $^{54}$Ca \cite{Steppenbeck2013}. However, most of these studies are based on phenomenological interactions where the nucleon-nucleon matrix elements were adjusted empirically. 
The spin-tensor analysis performed for effective interactions of microscopic origin in Ref. \cite{Smirnova2012} fell short in capturing accurate shell evolution, primarily due to the absence of three-body forces. The knowledge of different components of microscopic nucleon-nucleon interactions that reproduce observed shell closures and their impact on establishing magicity in exotic systems is still limited.

Recent advances in \textit{ab initio} theory have enabled the nonperturbative decoupling of effective interactions \cite{Ragnar2019} from realistic two- and three-nucleon potentials through various approaches, such as the no core shell model (NCSM) \cite{ncsm_pshell, ncsm_sdshell}, coupled cluster theory (CC) \cite{Jansen2014, Jansen2016}, and in medium similarity renormalization group (IMSRG) \cite{HergertReport, imsrg_bogner, imsrg_RagnarPRC, imsrg_RagnarPRL, Morris2018}. While NCSM is limited to lower mass nuclei \cite{Chandan2023, Na_work_NPA}, the applicability of CC \cite{HuPLB2024, HuPRCL2024} and IMSRG \cite{Miyagi2022PRC, Hu2022NatPhy, YuanPRCL2024, YuanN50PLB, TichaiN50PLB} methods has been extended to medium- and heavier-mass regions. With these developments, it is essential to address the evolution of shell structure from the underlying $2N$ and $3N$ forces. A thorough understanding of the microscopic origins behind the emergence of new magic numbers is needed. Moreover, with recent experimental advances, the study of exotic isotopes has become a central focus, making concurrent theoretical investigations into their structure increasingly important.

In this Letter, we study the evolution of the shell structure at $N=32$ and 34 using the state-of-the-art valence-space IMSRG (VS-IMSRG) method. Starting from QCD-based chiral $2N$ and $3N$ potentials, the VS-IMSRG method has demonstrated remarkable success in describing binding energies, low-lying spectroscopy, and electromagnetic properties of both closed- and open-shell nuclei \cite{imsrg_RagnarPRC, imsrg_RagnarPRL, Na_work_NPA, YuanPRCL2024}. The breaking of conventional shell gaps at $N=20$, 28 and the appearance of new magic numbers at $N=14$ are also well reproduced within the VS-IMSRG framework \cite{imsrg_Miyagi, OddNeMg_IoI, YuanN28PLB, Li2023}. In the present work, the VS-IMSRG interactions are employed to investigate the $N=32$ and 34 shell gaps in a series of isotopes and isotones near doubly magic $^{48}$Ca. Various facets of nuclear forces, such as central, spin-orbit, and tensor components, are addressed by spin-tensor decomposition of VS-IMSRG interactions. The roles played by these components in establishing shell gaps, particularly by the tensor components, are highlighted throughout the study. Then, we presented the low-lying structures of the exotic $N=32$ isotones below Ca, which are of current experimental interest \cite{Rosenbusch2015, Liu2019, Cortes2020, Linh2024}. Their low-energy structures, including spectroscopy of excited states, orbital occupations, and electromagnetic observables, are studied in detail. We begin with a brief overview of the VS-IMSRG formalism and outline the calculation details. Subsequently, the  \textit{ab initio} results and their interpretations are discussed, followed by a short summary of the work.

\textit{Method.} In the IMSRG approach \cite{Ragnar2019, HergertReport}, we start from the intrinsic $A$- body Hamiltonian given by
\begin{equation} \label{eq1}
    H = \frac{1}{A}\sum_{i<j}^A  \frac{{(\vec{p_i}-\vec{p_j})}^2}{2m} +
         \sum_{i<j}^A V_{ij}^{2N} + \sum_{i<j<k}^A V_{ijk}^{3N},
\end{equation}
where $\vec{p}$ is the nucleon momentum in the laboratory frame, $m$ represents the nucleon mass, and $V_{ij}^{2N}$ and $V_{ijk}^{3N}$ correspond to $2N$ and $3N$ nuclear forces, respectively. The well-established chiral EM1.8/2.0 interaction \cite{EMinteraction1, EMinteraction2}, comprising a next-to-next-to-next-to-leading order (N$^3$LO) $2N$ potential evolved via similarity renormalization group (SRG) to $\lambda$=1.8 fm$^{-1}$ and a $3N$ force at next-to-next-to-leading order (N$^2$LO) with momentum cutoﬀ $\Lambda$=2.0 fm$^{-1}$, is used for this purpose. In practice, the Hamiltonian is written in terms of creation ($a^{\dagger}_i$) and annihilation ($a_i$) operators as
\begin{align} \label{eq2}
    H=E_0+\sum_{ij}f_{ij}{ \lbrace a^\dagger_i a_j \rbrace }+\frac{1}{4}\sum_{ijkl}\Gamma_{ijkl}{ \lbrace a^\dagger_i a^\dagger_j a_l a_k \rbrace } \nonumber \\
    + \frac{1}{36}\sum_{ijklmn}W_{ijklmn}{ \lbrace a^\dagger_i a^\dagger_j a^\dagger_k  a_n a_m a_l \rbrace }.
\end{align}
Here $E_0$, $f$, $\Gamma$, and $W$ correspond to zero-, one-, two-, and three-body operators, respectively, and are normal ordered with respect to an ensemble reference state \cite{Ragnar2019, imsrg_RagnarPRL}. 

Within the VS-IMSRG framework, an effective Hamiltonian is decoupled from the large Hilbert space for a chosen valence space by applying a continuous unitary transformation $U(s)$. This is achieved by evolving the Hamiltonian through the SRG flow equation:
\begin{equation} \label{eq3}
\frac{dH(s)}{ds}=\left[ \eta(s),H(s)\right],
\end{equation}
where $s$ is called the flow parameter and $\eta(s)=\frac{dU(s)}{ds}U^{\dagger}(s)$ is known as the anti-Hermitian generator. The contributions of $3N$ forces are well captured up to the two-body level due to normal ordering \cite{Hebeler2023}. In practice, the Hamiltonian [Eq. \eqref{eq2}] is truncated at the two-body level in the IMSRG evolution to overcome computational and numerical challenges. This is known as the IMSRG(2) approximation, and the Magnus formalism \cite{MagnusIMSRG} is used to solve Eq. \eqref{eq3}.

The IMSRG calculations are carried out in a harmonic oscillator basis at $\hbar \omega$=16 MeV with $e=2n+l \leq e_{max}= 12$, and the additional truncation $e_1+e_2+e_3 \leq E_{3max}=24$ is imposed on $3N$ forces. The VS-IMSRG interactions are decoupled for the $fp$ shell for both protons and neutrons above the $^{40}$Ca core for nuclei with $Z \geq 20$. For neutron-rich nuclei with $Z<20$, our valence space spans protons in the $sd$ shell and neutrons in the $fp$ shell above a $^{28}$O core.
The effective Hamiltonians and consistently evolved effective $E2$ operators for the chosen valence space are generated through IMSRG calculations performed using the IMSRG++ \cite{imsrgCode} code of Ref. \cite{imsrg_RagnarPRL}. The valence-space Hamiltonians are diagonalized, and the corresponding transition densities are obtained through the KSHELL code \cite{kshell}.

\begin{figure} 
	\centering 
	\includegraphics[scale=0.54]{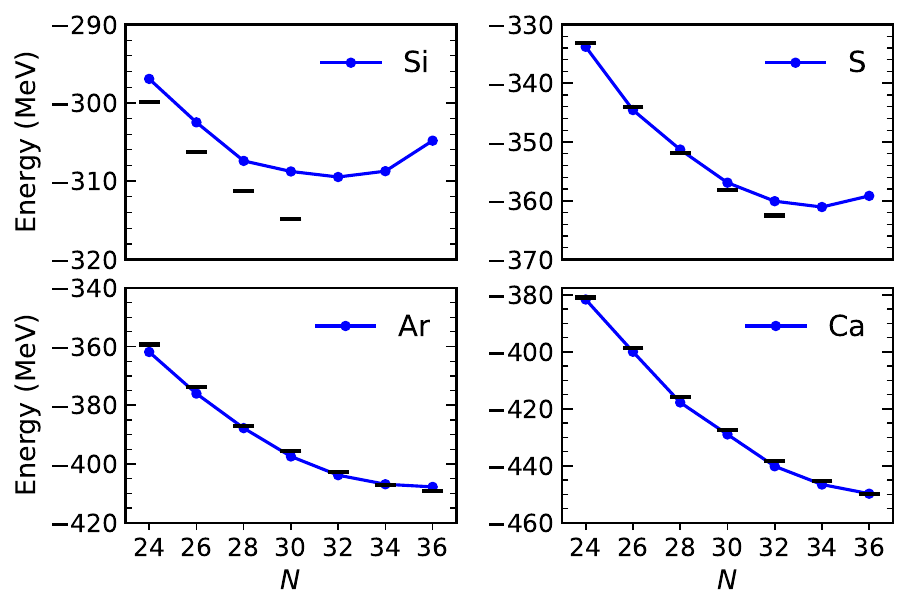}	
	\caption{Calculated (blue) and experimental (black) \cite{NNDC}  ground state binding energies of nuclei with $Z \leq 20$.} 
    \label{GSenergy}
\end{figure}

\textit{Results and discussion.} The $N=32$ and 34 subshell gaps are investigated in neutron-rich isotopes with $16 \leq Z \leq 26$. Since these isotopes, particularly those with $Z < 20$, lie on the extreme neutron-rich side and in medium mass regions, it is important to track the neutron drip line as well as the convergence of \textit{ab initio} calculations with $3N$ forces. To ensure convergence, in this work, the VS-IMSRG calculations are performed with $E_{3max}=24$, which has demonstrated reasonable convergence in heavier mass regions \cite{Miyagi2022PRC, Hu2022NatPhy}. With this truncation, the excitation energies of low-lying states in $^{60}$Fe remain almost identical from two sets of IMSRG calculations: $e_{max}=12$, $E_{3max}=22$ and $e_{max}=12$, $E_{3max}=24$, implying proper convergence. Figure \ref{GSenergy} shows the ground state (g.s.) binding energies for isotopic chains with $Z \leq 20$. The g.s. energies are well reproduced in the Ca, Ar, and S isotopic chains, while they are underestimated in neutron-rich Si isotopes. In the Si and S isotopic chains, the lowest g.s. energy is obtained at $N=32$ and $N=34$, respectively, after which it begins to increase with the addition of more neutrons. This suggests that the drip lines in the Si and S chains are reached at these neutron numbers, beyond which the two-neutron separation energy is negative, and additional isotopes are no longer bound. Therefore, we have confined our discussions on Si and S isotopes to their respective drip line nuclei.

\begin{figure*} 
	\centering
        \includegraphics[scale=0.70]{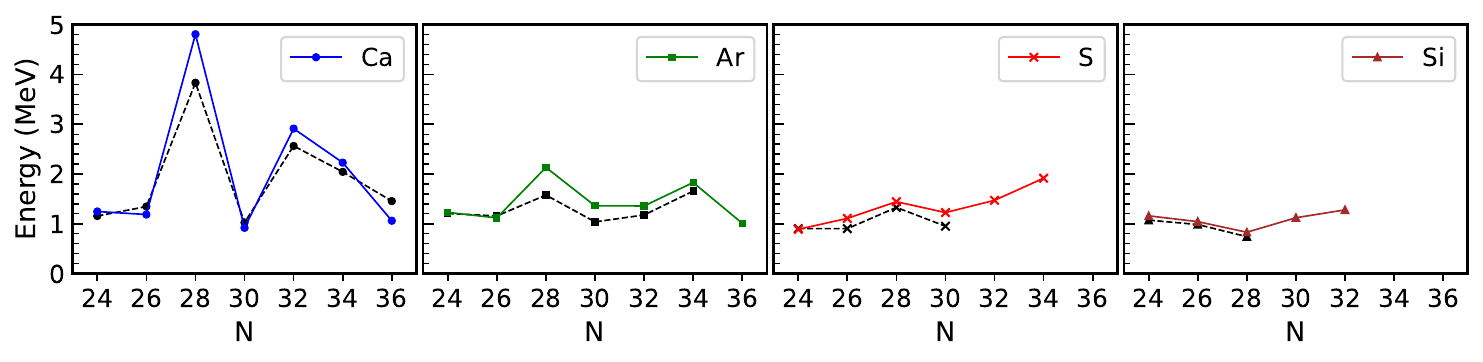}
        \includegraphics[scale=0.70]{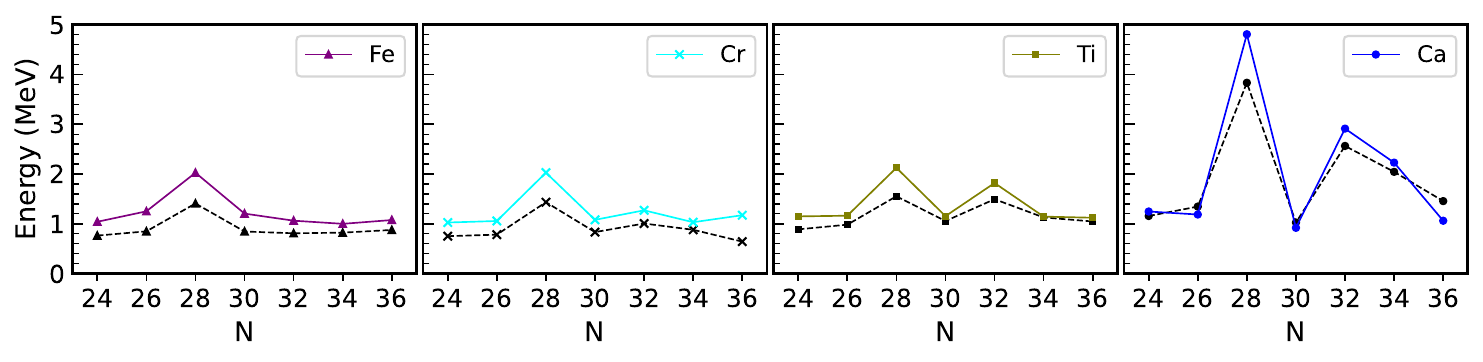}
	\caption{ Calculated (colored solid lines) and experimental \cite{NNDC} (black dotted lines) represent energies of the $2_1^+$ state in $26 \leq Z \leq 20$ (upper panel) and $20 \leq Z \leq 14$ (lower panel) isotopes.} 
    \label{ExStatesEnergy}
\end{figure*}

The energy trends of the first $2^+$ states [$E(2^+_1)$] in even-even nuclei along an isotopic chain are key indicators of shell or subshell closures. The calculated $E(2^+_1)$ values for $20 \leq Z \leq 26$ and $14 \leq Z \leq 20$ isotopes are compared with experimental data in the upper and lower panels of Fig. \ref{ExStatesEnergy}, respectively. These calculated values are generally higher than their experimental counterparts, likely due to the IMSRG(2) approximation, where induced three- and higher-body terms are neglected to reduce computational burdens. 
Recent studies incorporating IMSRG(3) calculations in small model spaces \cite{IMSRG3N7_Ragnar, IMSRG3N7_Heinz2024} at a higher computational cost and factorized approximations to IMSRG(3) \cite{IMSRG3f2} have led to improved agreement with experimental spectroscopic data. Nevertheless, the overall systematic trends of the $2^+_1$ states are well reproduced within VS-IMSRG calculations. In the upper panel of Fig. \ref{ExStatesEnergy}, the relatively high $E(2^+_1)$ at $N=32$ indicates the presence of a $N=32$ subshell gap in Ca and Ti, which weakens in Cr and gradually vanishes at Fe. The $E(2^+_1)$ trends show no signs of $N=34$ subshell closure in Fe to Ti, which instead appears only in the Ca chain, closely aligning with experimental observations.
In the lower panel of Fig. \ref{ExStatesEnergy}, the $E(2^+_1)$ systematics point towards the persistence of this $N=34$ subshell gap in Ar and S, while the $N=32$ subshell closure is weak or does not extend below Ca.

Effective (spherical) single-particle energies (ESPEs) of shell model orbits are useful theoretical tools to understand shell evolutions in nuclei, using both phenomenological \cite{OtsukaPRL2001, OtsukaPRL2005, SmirnovaPLB2010, OtsukaPRL2010, UtsunoReview2022} and realistic \cite{Smirnova2012, ncsm_sdshell, Ma2019} Hamiltonians. They correspond to the average effects of valence nucleons on the energies of single-particle orbitals. Assuming filling of nucleons in normal configuration, the ESPE ($\epsilon$) of an orbit $j$ is evaluated from the angular momentum averaged monopole components ($V_{jj'}^{mon.}$) of the effective Hamiltonian as follows:

\begin{equation}
    \textrm{ESPE}(\epsilon_j) = \epsilon_{0j} + \sum_{j'} V_{jj'}^{mon.} n_{j'}.
\end{equation}
Here $\epsilon_{0}$ is the bare single-particle energy, and the sum runs over valence-space orbitals $j'$, $n_{j'}$ being their occupation numbers. We adopt this formalism to calculate ESPEs and discuss the shell evolution from the energy gaps between relevant orbits of ESPEs. However, alternative approaches \cite{Duguet2012, Duguet2015, Soma2024} also exist that derive ESPEs in a more fundamental way within the context of strongly correlated many-nucleon systems.

\begin{figure}[h]
	\centering 
	\includegraphics[scale=0.62]{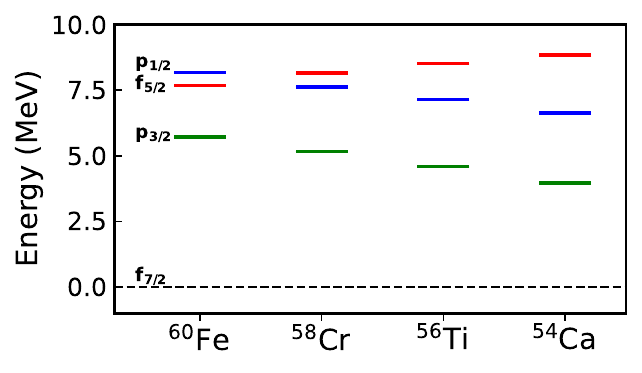}
	\caption{Neutron ESPEs of $N=34$ isotones in the $fp$  shell.} 
    \label{espes}
\end{figure}

Figure \ref{espes} shows the evolution of neutron ESPEs relative to the lowest $\nu f_{7/2}$ orbital in Fe to Ca. The ESPEs exhibit exactly similar patterns at both $N=32$ and $N=34$ with very small differences in their values, so we discuss the $N=32$ and 34 shell gaps from the ESPEs of $N=34$ isotones. In Fe, the $N=32$ subshell gap is disrupted by the presence of the $\nu f_{5/2}$ orbital in between $\nu p_{1/2}$ and $\nu p_{3/2}$. This $\nu f_{5/2}$ orbital rises above $\nu p_{1/2}$ in Cr and opens up the $N=32$ subshell gap. Moving to Ti and Ca, the $\nu p_{1/2}- \nu p_{3/2}$ energy gap increases moderately, reinforcing the $N=32$ subshell gap. Meanwhile,  the $\nu f_{5/2}$–$\nu p_{1/2}$ separation grows significantly from Cr to Ca, establishing the $N=34$ shell closure in Ca. The evolution of the $\nu f_{5/2}$ orbital is thus key to both subshell gaps. Otsuka \textit{et al.} \cite{OtsukaPRL2001} explained this through a strong tensor force between the spin-orbit partners $\pi f_{7/2}$ and $\nu f_{5/2}$, which weakens as protons are removed from $\pi f_{7/2}$ (Fe to Ca), causing the upward shift of $\nu f_{5/2}$ and facilitating the development of the $N=32$ and 34 shell gaps.

\begin{figure*}[htbp]
    \raisebox{0.35cm}{\includegraphics[width=0.32\textwidth]{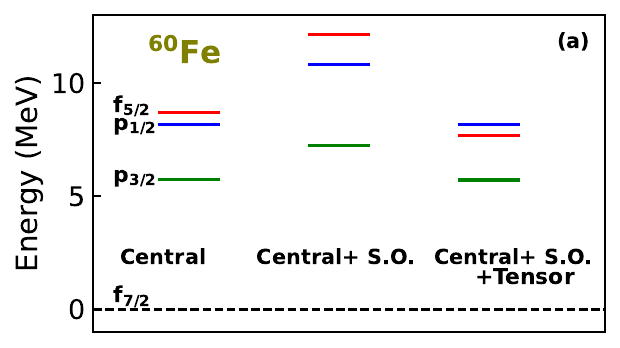}}
    \raisebox{0.00cm}{\includegraphics[width=0.32\textwidth]{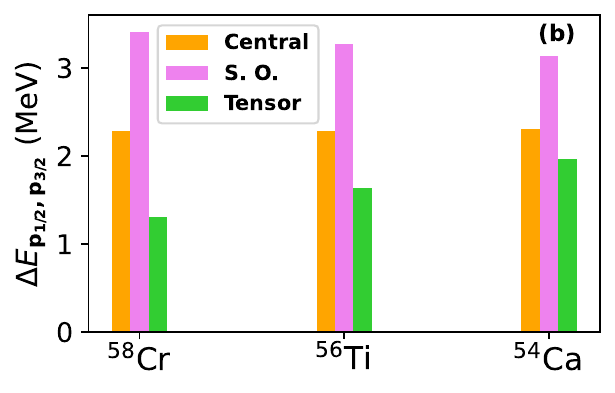}}
    \raisebox{0.00cm}{\includegraphics[width=0.32\textwidth]{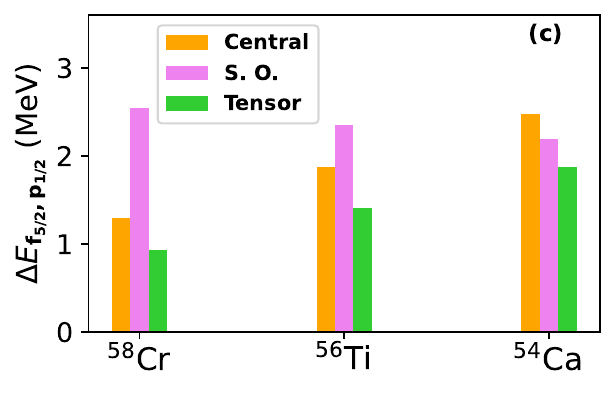}}
  \caption{(a) Central, spin-orbit (S.O.) and tensor contribution to neutron ESPEs in $^{60}$Fe. (b) $\nu p_{1/2} - \nu p_{3/2}$ energy gap from central, S.O., and tensor parts of ESPEs. (c) Same as (b), but for the $\nu f_{5/2} - \nu p_{1/2}$ gap.} \label{espe_components}
\end{figure*}

To understand the roles of different components of the nuclear force in the formation of shell gaps, we performed spin-tensor decomposition of the VS-IMSRG interactions following Refs. \cite{SmirnovaPLB2010, Kirson1973, Yoro1980, Brown1985}, where the two-body effective interactions or two-body matrix elements (tbmes) can be expressed as
\begin{equation}
    V = \sum_k U^{(k)}.C^{(k)} = \sum_{k=0,1,2} V^k.
\end{equation}
Here, $C^{(k)}$ and $U^{(k)}$ are rank-$k$ tensors in spin and configuration space, respectively, and $ V^k$ corresponds to the central ($k=0$), spin-orbit ($k=1$), and tensor ($k=2$) components of the effective interaction. These components are calculated from the tbmes in the $LS$ coupled scheme (see Supplemental Material \cite{SupplMat} for details) and subsequently converted back to the $jj$ scheme for further analysis. Then, their contributions to the development of shell gaps are analyzed through ESPEs, as shown in Fig. \ref{espe_components}.

As illustrated in Fig. \ref{espe_components}(a), the tensor component brings 
the $\nu f_{5/2}$ orbital to a position between $\nu p_{1/2}$ and $\nu p_{3/2}$ in Fe, thereby blocking the $N=32$ subshell gap. It weakens below Fe, and the $\nu f_{5/2}$ orbital appears above the $\nu p_{1/2}$-$\nu p_{3/2}$ energy gap from Cr to Ca, leading to the emergence of the $N=32$ subshell gap. However, the relative position of $\nu f_{5/2}$ orbital is determined by the combined effect of all three components. Figures \ref{espe_components}(b) and  \ref{espe_components}(c) present evolution of the $N=32$ ($\Delta E_{p_{1/2},p_{3/2}}$=$\nu p_{1/2}$-$\nu p_{3/2}$) and $N=34$ ($\Delta E_{f_{5/2},p_{1/2}}$=$\nu f_{5/2}$-$\nu p_{1/2}$) subshell gaps, obtained from the central, spin-orbit, and tensor components, as one progresses from Cr to Ca. The $N=32$ subshell gap remains almost constant in the central part, decreases slightly in the $k=1$ component, and is enhanced by the tensor component. The $k=1$ part has a similar complementary effect on the $N=34$ shell gap, while both central and tensor components collectively increase it. 
Thus, the central and tensor components emerge as leading contributors in shell evolution. 
While the $N=32$ shell gap is a direct consequence of the tensor force, the $N=34$ shell gap at Ca is shaped by the combined effects of central and tensor components.

\begin{figure}[h]
	\centering 
	\includegraphics[scale=0.59]{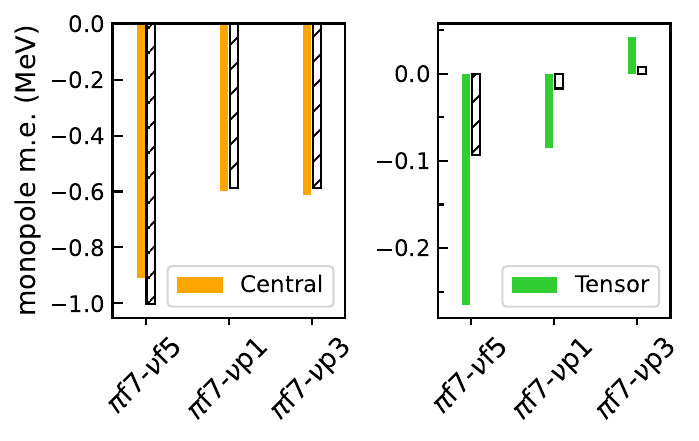}	
	\caption{Monopole m.e. of VS-IMSRG (solid bars) and GXPF1Br (hatched bars), with abbreviated orbital labels (e.g., f7 for $f_{7/2}$, etc.)} 
    \label{monopole_terms}
\end{figure}

The results are consistent with the findings observed from phenomenological models such as GXPF1Br \cite{Steppenbeck2013}, which reasonably describe the shell evolution in this region \cite{OtsukaReview}. To gain further insight, we compare the central and tensor proton-neutron monopole matrix elements (m.e.) of the VS-IMSRG and GXPF1Br interactions relevant to the $N=32$ and 34 shell gaps, as shown in Fig. \ref{monopole_terms}. The central part remains approximately similar in both cases, while the tensor terms exhibit noticeable differences. Within the VS-IMSRG framework, the tensor force, particularly between the spin-orbit partners $\pi f_{7/2}$-$\nu f_{5/2}$, which plays a key role in driving the $N=32$ and 34 shell gaps, is relatively stronger compared to phenomenological models.
Moreover, in some \textit{ab initio} approaches, the tensor components of the effective interaction exhibit renormalization persistency, closely resembling those of the bare nucleon-nucleon interaction \cite{tensor_rp1, tensor_rp2}. It is therefore of interest to explore in future studies whether this persistency holds within the VS-IMSRG framework.

\begin{figure*}
	\centering 
	\includegraphics[scale=0.69]{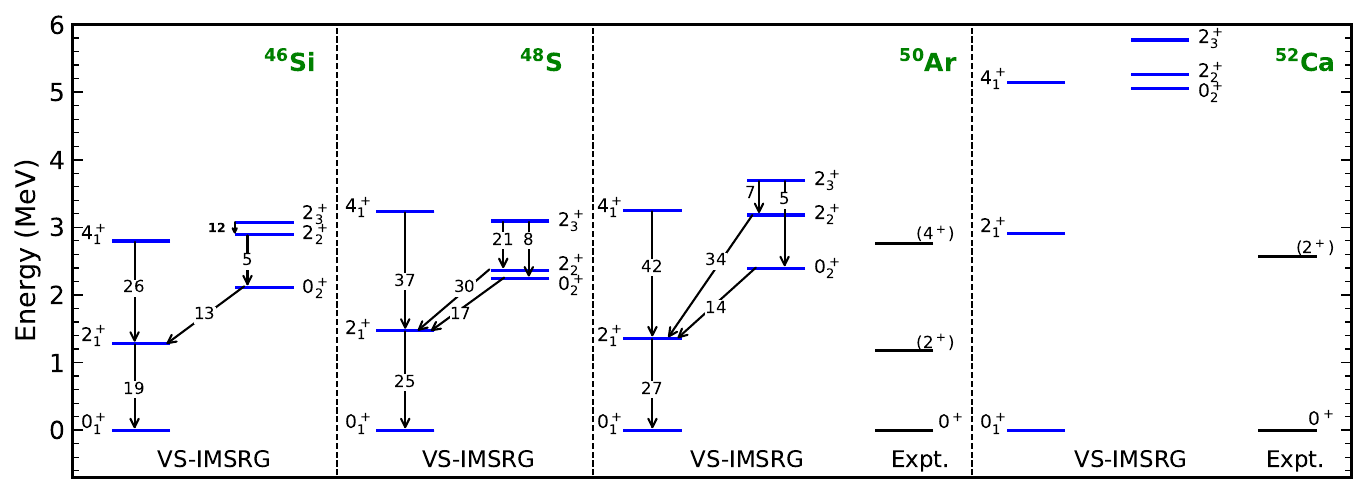}
	\caption{Calculated (blue) and experimental \cite{NNDC} (black) low-lying excitation spectrum and $B(E2)$ values (in e$^2$fm$^4$) in $N=32$ isotones with $Z \leq 20$. All $E2$ transitions between low-lying states with $B(E2) > 5$ e$^2$fm$^4$ are shown.} 
    \label{N32States}
\end{figure*}

Phenomenological studies \cite{Steppenbeck2015, Cortes2020, Linh2024} reported that the $N=32$ subshell gap in Ar is of similar magnitude to that observed in Ca, and the $\nu p_{1/2}-\nu p_{3/2}$ monopole gap changes only by $\approx 0.1$ MeV between them. However, the rise in $E(2_1^+)$ at $^{50}$Ar is notably less pronounced, and the ground state occupancy of the $\nu p_{1/2}$ orbital, calculated using these Hamiltonians, is considerably higher ($\approx 0.70$) than that found in $^{52}$Ca ($\approx 0.17$) \cite{Linh2024}. But, within VS-IMSRG interactions, the magnitude of the $\nu p_{1/2}-\nu p_{3/2}$ splitting decreases by $\approx 0.35$ MeV from Ca to Ar. The VS-IMSRG results suggest a weak $N=32$ subshell gap in Ar, comparable to that observed in Cr. Moreover, the VS-IMSRG predicts further reduction of the $N=32$ subshell gap in S, which gradually disappears in Si (see Ref. \cite{SupplMat}). But, the $N=34$ shell gap in $^{52}$Ar is comparatively larger (by $\approx 0.3$ MeV ) than that in $^{54}$Ca, and continues to widen in $^{50}$S and $^{46}$Si. Thus, while the $N=34$ shell gap becomes more pronounced in exotic isotopes below Ca, the strength of the $N=32$ subshell gap diminishes. The interplay among all three components ($k=0$, 1, and 2) enhances the $N=34$ shell gap and reduces the $N=32$ shell gap from Ar to Si.

\noindent
\textit{$N=32$ isotones.} Since the $N=32$ subshell effect is weak for $Z < 20$, it is worthwhile to discuss the structural characteristics of these isotones adjacent to the $N=34$ magic number. The computed low-energy excitation spectra of the $N=32$ isotones are shown in Fig. \ref{N32States} along with the $E2$ transition strengths connecting them. The VS-IMSRG results agree well with the experimental data for $^{50}$Ar and $^{52}$Ca. Meanwhile, the \textit{ab initio} predictions for $^{48}$S and $^{46}$Si offer valuable insights into the structure of these exotic isotopes in the absence of experimental data and can serve as a useful reference for future measurements.

The absolute $E2$ strengths predicted by VS-IMSRG are generally underestimated when compared to experimental values \cite{Henderson2018, imsrg_Miyagi}, likely due to the absence of higher-order collective excitations at the IMSRG(2) level, as discussed in Ref. \cite{E2_strength_vsimsrg}. However, these calculated $E2$ strengths reasonably reproduce the observed $B(E2)$ trends, and their relative values are physically meaningful \cite{YuanN50PLB, YuanN28PLB, YuanPRCL2024, imsrg_Miyagi}.
Based on the calculated excitation energies and interconnected $B(E2)$ values, two bands are predicted for the $N=32$ isotones from Ar to Si: a ground-state band connecting the $0_1^+$, $2_1^+$, and $4_1^+$ states, and a second low-lying band built on the $0_2^+$ state. This second band includes $0_2^+$ and $2_3^+$ states in $^{50}$Ar and $^{48}$S, while in $^{46}$Si the $2_2^+$ appears as its band member instead of $2_3^+$.
The relatively strong interband $E2$ transitions indicate configuration mixing. In contrast, $^{52}$Ca exhibits a closed shell configuration, as depicted in Fig. \ref{N32States}. The VS-IMSRG results are consistent with the experimental observations and shell model predictions, showing the onset of collectivity in $^{50}$Ar \cite{Linh2024}. Our results suggest that this collective behavior continues further in S and Si.

\begin{table}[h]
    \centering
    \caption{Calculated quadrupole moments of the $2^+$ states and occupation numbers of the $\nu p_{1/2}$ and $\nu f_{5/2}$ orbitals for $0_1^+$ and $0_2^+$ states in $N=32$ isotones.}
    \label{Table2}
    \setlength{\tabcolsep}{6pt} 
    \renewcommand{\arraystretch}{1.2} 
    \begin{tabular}{l cc cc cc}
        \hline \hline
        Nuclei & \multicolumn{2}{c}{$Q_s$ (efm$^2$)} & \multicolumn{2}{c}{$n_{\nu}[p_{1/2}]$} & \multicolumn{2}{c}{$n_{\nu}[f_{5/2}]$} \\
        \cline{2-3} \cline{4-5} \cline{6-7}
        & $2_1^+$ & $2_3^+$ & $0_1^+$ & $0_2^+$ & $0_1^+$ & $0_2^+$ \\
        \hline
        $^{50}$Ar & 10.8 & 1.6  & 1.005 & 1.013 & 0.408 & 0.456 \\
        $^{48}$S  & 6.1  & -2.5 & 1.059 & 0.920 & 0.637 & 0.724 \\
        $^{46}$Si & 8.5  & -1.4\footnotemark[1] & 1.292 & 0.790 & 0.657 & 0.627 \\
        \hline
    \end{tabular}
    \footnotetext[1]{Corresponds to $2_2^+$ instead of $2_3^+$.}
\end{table}

The spectroscopic quadrupole moments of the $2^+$ states [$Q_s(2^+)$], associated with different bands of $^{50}$Ar, $^{48}$S, and $^{46}$Si, are listed in Table \ref{Table2}. The $Q_s(2_2^+ / 2_3^+)$ values are small whereas the $Q_s(2_1^+)$ values are notably larger and positive. This implies that the ground state is strongly oblate deformed and coexists with a weakly deformed excited band. Table \ref{Table2} also shows the neutron occupancies ($n_{\nu}$) of $p_{1/2}$ and $f_{5/2}$ orbitals ($n_{\nu}$[$p_{1/2}$] and $n_{\nu}$[$f_{5/2}]$) for $0_1^+$ and $0_2^+$ states, which lie beyond the normal filling of $N=32$ isotones. Nearly identical occupancies in both states suggest similar cross-shell configurations and account for the strong $E2$ transition between the two bands. In the ground states, $n_{\nu}$[$p_{1/2}]$ is approximately 1.1, indicating quenching of the $N=32$ shell gap. In contrast, this value is much lower in the doubly magic $^{52}$Ca ($\approx 0.16$). The $n_{\nu}$[$f_{5/2}]$ values also remains small ($\approx 0.7$) in $^{52}$Ar and $^{50}$S, reflecting the prominence of the $N=34$ shell gap. Earlier, VS-IMSRG calculations suggested the dominance of cross-shell configurations and deformed $^{42}$Si and $^{44}$S nuclei at $N=28$ \cite{YuanN28PLB}. As one progresses to more exotic isotopes, spherical configurations emerge at $N=34$.

\textit{Summary.} The shell evolution at $N=32$ and $N=34$ in neutron-rich isotopes with $14 \leq Z \leq 26$ is studied from chiral $2N$ and $3N$ forces within the VS-IMSRG framework. The calculated results are in good agreement with the experimental data. Within the $fp$ shell, the VS-IMSRG results suggest $N=32$ subshell closure in Ca, Ti, which weakens in Cr and gradually disappears in Fe. The $N=34$ subshell gap emerges exclusively in Ca. Remarkably, this subshell gap continues to persist even below Ca, extending as far as Si, while the $N=32$ subshell is weak or vanishes below Ca. Through spin-tensor decomposition, the roles of different components of the nuclear force--central, spin-orbit, and tensor--in the development of shell gaps are investigated.
Next, we have discussed the low-lying structures of the exotic $N=32$ isotones for $Z \leq 20$ adjacent to the $N=34$ magic number. The VS-IMSRG calculations predict the presence of intruder configurations and large oblate deformation in the ground state band of $^{50}$Ar, $^{48}$S, and $^{46}$Si isotopes, coexisting with a weakly deformed excited band. These \textit{ab initio} results provide deeper insights into the structure of exotic nuclei, highlighting various components of realistic nuclear forces and their roles in establishing magic numbers far from stability.

\textit{Acknowledgements:}
We are grateful to Takayuki Miyagi for availing the chiral $NN$ and $NNN$ nuclear matrix elements, generated by the NuHamil code \cite{NuHamil}. S.S. would like to thank UGC (University Grant Commission), India, for financial support for his Ph.D. thesis work. P.C.S. acknowledges a research grant from SERB (India), CRG/2022/005167. We would like to thank the National Supercomputing Mission (NSM) for providing computing resources of ‘PARAM Ganga’ at the Indian Institute of Technology Roorkee, implemented by C-DAC and supported by the Ministry of Electronics and Information Technology (MeitY) and Department of Science and Technology (DST), Government of India.

\textit{Data availability}. The data supporting this study’s findings
are available within the article.



\begin{thebibliography}{44}

\bibitem{SorlinReview}
O. Sorlin and M.-G. Porquet,
\href{http://dx.doi.org/10.1016/j.ppnp.2008.05.0010}
{Prog. Part. Nucl. Phys. {\bf 61}, 602 (2008).}

\bibitem{OtsukaReview}
T. Otsuka, A. Gade, O. Sorlin, T. Suzuki, and Y. Utsuno,
\href{https://doi.org/10.1103/RevModPhys.92.015002}
{Rev. Mod. Phys. {\bf 92}, 015002 (2020).}

\bibitem{Hagen2012}
G. Hagen, M. Hjorth-Jensen, G. R. Jansen, R. Machleidt, and T. Papenbrock,
\href{https://doi.org/10.1103/PhysRevLett.109.032502}
{Phys. Rev. Lett. {\bf 109}, 032502 (2012).}

\bibitem{Hergert2014}
H. Hergert, S. K. Bogner, T. D. Morris, S. Binder, A. Calci, J. Langhammer, and R. Roth,
\href{https://doi.org/10.1103/PhysRevC.90.041302}
{Phys. Rev. C {\bf 90}, 041302(R) (2014).}

\bibitem{Soma2021}
V. Somà, C. Barbieri, T. Duguet, and P. Navrátil,
\href{https://doi.org/10.1140/epja/s10050-021-00437-4}
{Eur. Phys. J. A {\bf 57}, 135 (2021).}




\bibitem{Steppenbeck2013}
D. Steppenbeck \textit{et al.},
\href{https://doi.org/10.1038/nature12522}
{Nature {\bf 502}, 207 (2013).}

\bibitem{Xu2019}
X. Xu \textit{et al.},
\href{http://dx.doi.org/10.1103/PhysRevC.99.064303}
{Phys. Rev. C {\bf 99}, 064303 (2019).}

\bibitem{Leistenschneider2021}
E. Leistenschneider \textit{et al.},
\href{http://dx.doi.org/10.1103/PhysRevLett.126.042501}
{Phys. Rev. Lett. {\bf 126}, 042501 (2021).}




\bibitem{Iimura2023}
S. Iimura \textit{et al.},
\href{http://dx.doi.org/10.1103/PhysRevLett.130.012501}
{Phys. Rev. Lett. {\bf 130}, 012501 (2023).}

\bibitem{Steppenbeck2015}
D. Steppenbeck \textit{et al.},
\href{https://doi.org/10.1103/PhysRevLett.114.252501}
{Phys. Rev. Lett. {\bf 114}, 252501 (2015).}

\bibitem{Rosenbusch2015}
M. Rosenbusch \textit{et al.},
\href{https://doi.org/10.1103/PhysRevLett.114.202501}
{Phys. Rev. Lett. {\bf 114}, 202501 (2015).}

\bibitem{Liu2019}
H. N. Liu \textit{et al.},
\href{https://doi.org/10.1103/PhysRevLett.122.072502}
{Phys. Rev. Lett. {\bf 122}, 072502 (2019).}

\bibitem{Cortes2020}
M. L. Cortés \textit{et al.},
\href{https://doi.org/10.1103/PhysRevLett.114.252501}
{Phys. Rev. C {\bf 102}, 064320 (2020).}

\bibitem{Linh2024}
B. D. Linh \textit{et al.},
\href{https://doi.org/10.1103/PhysRevC.109.034312}
{Phys. Rev. C {\bf 109}, 034312 (2024).}

 

\bibitem{OtsukaPRL2001}
T. Otsuka, R. Fujimoto, Y. Utsuno, B. A. Brown, M. Honma, and T. Mizusaki,
\href{https://doi.org/10.1103/PhysRevLett.87.082502}
{Phys. Rev. Lett. {\bf 87}, 082502 (2001).}

\bibitem{OtsukaPRL2005}
T. Otsuka, T. Suzuki, R. Fujimoto, H. Grawe, and Y. Akaishi,
\href{https://doi.org/10.1103/PhysRevLett.95.232502}
{Phys. Rev. Lett. {\bf 95}, 232502 (2005).}

\bibitem{Honma2002}
M. Honma, T. Otsuka, B. A. Brown, and T. Mizusaki,
\href{https://doi.org/10.1103/PhysRevC.65.061301}
{Phys. Rev. C {\bf 65}, 061301(R) (2002).}

\bibitem{Honma2004}
M. Honma, T. Otsuka, B. A. Brown, and T. Mizusaki,
\href{https://doi.org/10.1103/PhysRevC.69.034335}
{Phys. Rev. C {\bf 69}, 034335 (2004).}

\bibitem{Honma2005}
M. Honma, T. Otsuka, B. A. Brown, and T. Mizusaki,
\href{https://doi.org/10.1140/epjad/i2005-06-032-2}
{Eur. Phys. J. A {\bf 25}, 499 (2005).}


\bibitem{OtsukaPRL2010}
T. Otsuka, T. Suzuki, M. Honma, Y. Utsuno, N. Tsunoda, K. Tsukiyama, and M. H. Jensen,
\href{https://doi.org/10.1103/PhysRevLett.104.012501}
{Phys. Rev. Lett. {\bf 104}, 012501 (2010).}

\bibitem{BhoyCa2020}
B. Bhoy, P. C. Srivastava and K. Kaneko,
\href{https://doi.org/10.1088/1361-6471/ab80d4}
{J. Phys. G: Nucl. Part. Phys. {\bf 47}, 065105 (2020).}

\bibitem{UtsunoReview2022}
Y. Utsuno,
\href{https://doi.org/10.3390/physics4010014}
{Physics {\bf 4}, 185 (2022).}



\bibitem{Smirnova2012}
N. A. Smirnova, K. Heyde, B. Bally, F. Nowacki, and K. Sieja,
\href{https://doi.org/10.1103/PhysRevC.86.034314}
{Phys. Rev. C {\bf 86}, 034314 (2012).}




\bibitem{Ragnar2019}
S. R. Stroberg, H. Hergert, S. K. Bogner, and J. D. Holt, 
\href{https://doi.org/10.1146/annurev-nucl-101917-021120}
{Annu. Rev. Nucl. Part. Sci. {\bf 69}, 307 (2019).}


\bibitem{ncsm_pshell}
E. Dikmen, A. F. Lisetskiy, B. R. Barrett, P. Maris, A. M. Shirokov and J. P. Vary,
\href{http://dx.doi.org/10.1103/PhysRevC.91.064301}
{Phys. Rev. C {\bf 91}, 064301 (2015).}

\bibitem{ncsm_sdshell}
N. A. Smirnova, B. R. Barrett, Y. Kim, I. J. Shin, A. M. Shirokov, E. Dikmen, P. Maris and J. P. Vary,
\href{https://doi.org/10.1103/PhysRevC.100.054329}
{Phys. Rev. C {\bf 100}, 054329 (2019).}


\bibitem{Jansen2014}
G. R. Jansen, J. Engel, G. Hagen, P. Navrátil, and A. Signoracci,
\href{http://dx.doi.org/10.1103/PhysRevLett.113.142502}
{Phys. Rev. Lett. {\bf 113}, 142502 (2014).}


\bibitem{Jansen2016}
G. R. Jansen, M. D. Schuster, A. Signoracci, G. Hagen, and P. Navrátil,
\href{http://dx.doi.org/10.1103/PhysRevC.94.011301}
{Phys. Rev. C. {\bf 94}, 011301(R) (2016).}


\bibitem{HergertReport}
H. Hergert, S.K. Bogner, T.D. Morris, A. Schwenk, K. Tsukiyama,
\href{http://dx.doi.org/10.1016/j.physrep.2015.12.007}
{Phys. Rep. {\bf 621}, 165 (2016).}

\bibitem{imsrg_bogner}
S. K. Bogner, H. Hergert, J. D. Holt, A. Schwenk, S. Binder, A. Calci, J. Langhammer, and R. Roth,
\href{http://dx.doi.org/10.1103/PhysRevLett.113.142501}
{Phys. Rev. Lett. {\bf 113}, 142501 (2014).}


\bibitem{imsrg_RagnarPRC}
S. R. Stroberg, H. Hergert, J. D. Holt, S. K. Bogner and A. Schwenk,
\href{http://dx.doi.org/10.1103/PhysRevC.93.051301}
{Phys. Rev. C {\bf 93}, 051301(R) (2016).}


\bibitem{imsrg_RagnarPRL}
S. R. Stroberg, A. Calci, H. Hergert, J. D. Holt, S. K. Bogner, R. Roth and A. Schwenk,
\href{http://dx.doi.org/10.1103/PhysRevLett.118.032502}
{Phys. Rev. Lett. {\bf 118}, 032502 (2017).}

\bibitem{Morris2018}
T. D. Morris, J. Simonis, S. R. Stroberg, C. Stumpf, G. Hagen, J. D. Holt, G. R. Jansen, T. Papenbrock, R. Roth, and A. Schwenk,
\href{https://doi.org/10.1103/PhysRevLett.120.152503}
{Phys. Rev. Lett. {\bf 120}, 152503 (2018).}


\bibitem{Chandan2023}
C. Sarma and P. C. Srivastava,
\href{https://dx.doi.org/10.1088/1361-6471/acb962}
{J. Phys. G {\bf 50}, 045105 (2023)}.

\bibitem{Na_work_NPA}
S. Sahoo, P. C. Srivastava, and T. Suzuki,
\href{https://doi.org/10.1016/j.nuclphysa.2023.122618}
{Nucl. Phys. A {\bf 1032}, 122618 (2023)}.

\bibitem{HuPLB2024}
B. S. Hu, Z. H. Sun, G. Hagen, G. R. Jansen, T. Papenbrock,
\href{https://doi.org/10.1016/j.physletb.2024.139010}
{Phys. Lett. B {\bf 858}, 139010 (2024).}

\bibitem{HuPRCL2024}
B. S. Hu, Z.H. Sun, G. Hagen, and T. Papenbrock,
\href{https://doi.org/10.1103/PhysRevC.110.L011302}
{Phys. Rev. C {\bf 110}, L011302 (2024).}

\bibitem{Miyagi2022PRC}
T. Miyagi, S. R. Stroberg, P. Navrátil, K. Hebeler, and J. D. Holt,
\href{https://doi.org/10.1103/PhysRevC.105.014302}
{Phys. Rev. C {\bf 105}, 014302 (2022).}

\bibitem{Hu2022NatPhy}
B. Hu, W. Jiang, T. Miyagi, Z. Sun, A. Ekström, C. Forssén, G. Hagen, J. D. Holt, T. Papenbrock, S. R. Stroberg and I. Vernon,
\href{https://doi.org/10.1038/s41567-022-01715-8}
{Nat. Phys. {\bf 18}, 1196 (2022).}

\bibitem{TichaiN50PLB}
A. Tichai, K. Kapás, T. Miyagi, M.A. Werner, Ö. Legeza, A. Schwenk, and
G. Zarand,
\href{https://doi.org/10.1016/j.physletb.2024.138841}
{Phys. Lett. B {\bf 855}, 138841 (2024).}


\bibitem{YuanPRCL2024}
Q. Yuan, J. G. Li, and W. Zuo,
\href{https://doi.org/10.1103/PhysRevC.109.L041301}
{Phys. Rev. C {\bf 109}, L041301 (2024).}

\bibitem{YuanN50PLB}
Q. Yuan and B.S. Hu,
\href{https://doi.org/10.1016/j.physletb.2024.139018}
{Phys. Lett. B {\bf 858}, 139018 (2024).}


\bibitem{imsrg_Miyagi}
T. Miyagi, S. R. Stroberg, J. D. Holt and N. Shimizu,
\href{http://dx.doi.org/10.1103/PhysRevC.102.034320}
{Phys. Rev. C {\bf 102}, 034320 (2020).}

\bibitem{OddNeMg_IoI}
S. Sahoo and P.C. Srivastava,
\href{https://doi.org/10.1103/PhysRevC.111.054308}
{Phys. Rev. C {\bf 111}, 054308 (2025).}

\bibitem{YuanN28PLB}
Q. Yuan, J.G. Li, and H.H. Li,
\href{https://doi.org/10.1016/j.physletb.2023.138331}
{Phys. Lett. B {\bf 848}, 138331 (2024).}

\bibitem{Li2023}
J.G. Li, H.H. Li, S. Zhang, Y.M. Xing, and W. Zuo,
\href{https://doi.org/10.1016/j.physletb.2023.138197}
{Phys. Lett. B {\bf 846}, 138197 (2023).}

\bibitem{EMinteraction1}
K. Hebeler, S. K. Bogner, R. J. Furnstahl, A. Nogga, and A. Schwenk,
\href{https://doi.org/10.1103/PhysRevC.83.031301}
{Phys. Rev. C {\bf 83}, 031301(R) (2011).}

\bibitem{EMinteraction2}
J. Simonis, K. Hebeler, J. D. Holt, J. Men\'{e}ndez, and A.
Schwenk,
\href{http://dx.doi.org/10.1103/PhysRevC.93.011302}
{Phys. Rev. C {\bf 93}, 011302(R) (2016).}

\bibitem{Hebeler2023}
K. Hebeler, V. Durant, J. Hoppe, M. Heinz, A. Schwenk, J. Simonis, and A. Tichai,
\href{https://doi.org/10.1103/PhysRevC.107.024310}
{Phys. Rev. C {\bf 107}, 024310 (2023).}

\bibitem{MagnusIMSRG}
T. D. Morris, N. M. Parzuchowski, and S. K. Bogner,
\href{http://dx.doi.org/10.1103/PhysRevC.92.034331}
{Phys. Rev. C {\bf 92}, 034331 (2015).}


\bibitem{imsrgCode}
S. R. Stroberg,
\href{https://github.com/ragnarstroberg/imsrg}
{https://github.com/ragnarstroberg/imsrg.}

\bibitem{kshell} N. Shimizu, T. Mizusaki, Y. Utsuno and Y. Tsunoda,
\href{https://doi.org/10.1016/j.cpc.2019.06.011}
{Comput. Phys. Comm. {\bf 244}, 372 (2019)}.

\bibitem{IMSRG3N7_Ragnar}
S. R. Stroberg, T. D. Morris, and B. C. He,
\href{https://doi.org/10.1103/PhysRevC.110.044316}
{Phys. Rev. C {\bf 110}, 044316 (2024).}

\bibitem{IMSRG3N7_Heinz2024}
M. Heinz, T. Miyagi, S. R. Stroberg, A. Tichai, K. Hebeler, and A. Schwenk,
\href{https://doi.org/10.1103/PhysRevC.111.034311}
{Phys. Rev. C {\bf 111}, 034311 (2025).}

\bibitem{IMSRG3f2}
B. C. He and S. R. Stroberg,
\href{https://doi.org/10.1103/PhysRevC.110.044317}
{Phys. Rev. C {\bf 110}, 044317 (2024).}


\bibitem{NNDC}Data extracted using the National Nuclear Data Center World Wide Web site from the evaluated nuclear structure data file, 
\href{https://www.nndc.bnl.gov/ensdf/.}
{ https://www.nndc.bnl.gov/ensdf/}.


\bibitem{SmirnovaPLB2010}
N. A. Smirnova, B. Bally, K. Heyde, F. Nowacki, K. Sieja,
\href{https://doi.org/10.1016/j.physletb.2010.02.051}
{Phys. Lett. B {\bf 686}, 109 (2010).}

\bibitem{Ma2019}
Y. Z. Ma, L. Coraggio, L. De Angelis, T. Fukui, A. Gargano, N. Itaco, and F. R. Xu,
\href{https://doi.org/10.1103/PhysRevC.100.034324}
{Phys. Rev. C {\bf 100}, 034324 (2019).}

\bibitem{Duguet2012}
T. Duguet, and G. Hagen,
\href{http://dx.doi.org/10.1103/PhysRevC.85.034330}
{Phys. Rev. C {\bf 85}, 034330 (2012).}

\bibitem{Duguet2015}
T. Duguet, H. Hergert, J. D. Holt, and V. Soma,
\href{http://dx.doi.org/10.1103/PhysRevC.92.034313}
{Phys. Rev. C {\bf 92}, 034313 (2015).}

\bibitem{Soma2024}
V. Somà and T. Duguet,
\href{https://doi.org/10.1098/rsta.2023.0117}
{Phil. Trans. R. Soc. A {\bf 382}, 20230117 (2024).}

\bibitem{Kirson1973}
M. W. Kirson,
\href{https://doi.org/10.1016/0370-2693(73)90582-0}
{Phys. Lett. B {\bf 51}, 110 (1973).}

\bibitem{Yoro1980}
K. Yoro,
\href{https://doi.org/10.1016/0375-9474(80)90016-0}
{Nucl. Phys. A {\bf 333}, 67 (1980).}

\bibitem{Brown1985}
B. A. Brown, W. A. Richter and B. H. Wildenthalt,
\href{https://doi.org/10.1088/0305-4616/11/11/005}
{J. Phys. G {\bf 11}, 1191 (1985).}

\bibitem{SupplMat}
See Supplemental Material at 
\href{http://link.aps.org/supplemental/10.1103/423y-znv8}
{http://link.aps.org/supplemental/10.1103/423y-znv8}
10.1103/423y-znv8 for the spin-tensor decomposition method
and evolution of neutron ESPEs in exotic isotopes with
Z $<$ 20.

\bibitem{tensor_rp1}
N. Tsunoda, T. Otsuka, K. Tsukiyama, M. H. Jensen
\href{http://dx.doi.org/10.1103/PhysRevC.84.044322}
{Phys. Rev. C {\bf 84}, 044322 (2011).}

\bibitem{tensor_rp2}
N. Tsunoda, T. Otsuka, K. Tsukiyama, M. H. Jensen,
\href{https://doi.org/10.1088/1742-6596/267/1/012020}
{J. Phys. Conf. Series {\bf 267}, 012020 (2011).}

\bibitem{Henderson2018}
J. Henderson \textit{et al.},
\href{https://doi.org/10.1016/j.physletb.2018.05.064}
{Phys. Lett. B {\bf 782}, 468 (2018).}


\bibitem{E2_strength_vsimsrg}
S. R. Stroberg, J. Henderson, G. Hackman, P. Ruotsalainen, G. Hagen, and J. D. Holt,
\href{https://doi.org/10.1103/PhysRevC.105.034333}
{Phys. Rev. C {\bf 105}, 034333 (2022).}

\bibitem{NuHamil}
T. Miyagi,
\href{https://doi.org/10.1140/epja/s10050-023-01039-y}
{Eur. Phys. J. A {\bf 59}, 150 (2023).}



\end{thebibliography}
\end{document}